\documentclass[a4paper,twoside]{article}
\newcommand{\affil}[1]{$^{\rm #1}$}
\usepackage[authoryear]{natbib}
\bibpunct{(}{)}{;}{a}{}{,}
\usepackage{graphicx}
\usepackage{natbib}
\date{} 
\newcommand{\ms}{$\,$M$_\mathrm{\odot}$}

\newcommand{\be}{\begin{equation}}
\newcommand{\ee}{\end{equation}}
\newcommand{\stars}{{\sc stars}}
\newcommand{\el}[2]{\ensuremath{^{#1}\mathrm{#2}}}

\title{\large\bf\flushleft Why Do Low-Mass Stars Become Red Giants?}
\author{\parbox{\textwidth}{\flushleft
\vspace{-0.5cm}
{\it Richard J. Stancliffe\affil{A,C}, Alessandro Chieffi\affil{A,B}, John C. Lattanzio\affil{A}, and Ross P. Church\affil{A}}\\
\vspace{0.4cm}
{\small \affil{A}\,Centre for Stellar and Planetary Astrophysics, Monash University, PO Box 28M, Victoria 3800, Australia}\\
{\small \affil{B}\,INAF--Istituto di Astrofisica, Spaziale e Fisica Cosmica, Via Fosso del Cavaliere, Rome, Italy}\\
{\small \affil{C}\,Email: Richard.Stancliffe@sci.monash.edu.au}}}
\begin{document}
\twocolumn[
\begin{changemargin}{.8cm}{.5cm}
\begin{minipage}{.9\textwidth}
\vspace{-1cm}
\maketitle
%
%
\small{\bf Abstract:} We revisit the problem of why stars become red giants. We modify the physics of a standard stellar evolution code in order to determine what does and what does not contribute to a star becoming a red giant. In particular, we have run tests to try to separate the effects of changes in the mean molecular weight and in the energy generation. The implications for why stars become red giants are discussed. We find that while a change in the mean molecular weight is necessary (but not sufficient) for a 1\ms\ star to become a red giant, this is not the case in a star of 5\ms. It therefore seems that there may be more than one way to make a giant.


\medskip
\medskip
\end{minipage}
\end{changemargin}
]
\small

\section{Introduction}

The question `why do stars become red giants?' is perhaps one of the longest standing problems in stellar astrophysics. In a recent paper \citet{2000ApJ...538..837S} provided a long list of work on the subject, with publication dates spanning over four decades. The problem has been approached from many different angles, from considerations of polytropic solutions \citep[e.g.][]{1991ApJ...383..757E} to detailed numerical modelling \citep[e.g][]{1993ApJ...415..767I}. Despite all the investigation into the subject, the question has yet to receive an answer that is satisfyingly simple and sufficiently rigourous\footnote{Of course, we are labouring under the assumption that such a solution does indeed exist.}. There is still no consensus on why stars become red giants. Theories to explain the phenomenon are many and varied. Some assert that a `softening' of the effective equation of state is required \citep{1991ApJ...383..757E,1998MNRAS.298..831E}, others suggest that a strengthening of the central gravitational field is required \citep{1973A&A....25...99H, 1983A&A...127..411W}. Some attribute the behaviour to a gravothermal instability in the core \citep{1993ApJ...415..767I}, others suggest it is a thermal instability in the stellar envelope \citep{1992ApJ...400..280R}. The list of potential explanations is long and to review them all would take more pages than the authors have been permitted for this contribution. \citet{2000ApJ...538..837S} give a review (as well as a rather fierce rebuttal) of the leading theories and the interested reader is directed to this paper for a more thorough discussion.

Here we shall only briefly review some of the work that has been done on the problem via direct numerical experiments. \citet{1973A&A....25...99H} investigated the effects of an increased gravitational field in models of 1.4 and 1.8\ms, as well as looking at the effects of inhomogeneities in the mean molecular weight and in energy generation. They concluded that only a strong gravitational field could produce properties similar to red giants. Subsequently, \citet{1983A&A...127..411W} extended their work to cover stars in the mass range $1\leq M/\mathrm{M_\odot}\leq8$. Using polytropic models,  \citet{1992PASAu..10..125F} later demonstrated that this could not be the sole cause.

\citet{1992ApJ...400..280R} suggested that stars become red giants because of a thermal instability in their envelopes. In their view, expansion is initially driven by the envelope maintaining thermal equilibrium in response to increasing luminosity from the core. This expansion leads to local cooling and the recombination of heavy elements. An increase in the opacity results, trapping energy in the envelope. This leads to a runaway expansion that brings the star to the red giant branch. However, \citet{1993ApJ...415..767I} subsequently showed that the opacity could not be responsible for the transition to a red giant structure. He did this by computing models in which the opacity was held constant throughout the star. These models still became giants. 

In reviewing the literature, what is perhaps most striking is how contradictory much of the work is. For example, numerical experiments by \citet{1993ApJ...415..767I} show that the opacity does not play a key role in the process, yet \citet{2000ApJ...538..837S} assert that ``A key role is played by the gradients of opacity". One of the reasons for this may be that most studies are limited in their scope, looking only at models of a particular mass. Studies examining a range of masses are the exception, rather than the rule. A systematic study is clearly warranted.

In this work, we adopt the commonly-used approach of running experiments using a detailed evolution code. By altering the input physics in a controlled way and seeing what effect this has on the evolution, we hope to gain some understanding of what factors control whether a star becomes a red giant or not. We emphasise that by the phrase ``becomes a red giant'' we mean that the star develops a condensed core with an extended envelope, moving to cooler effective temperatures as it does so.

\section{The stellar evolution code}
Our calculations have been carried out using the \stars\ stellar evolution code originally developed by \citet{1971MNRAS.151..351E} and updated by many authors \citep[e.g.][]{1995MNRAS.274..964P}. The code follows the evolution of seven energetically important species, namely \el{1}{H}, \el{3}{He}, \el{4}{He}, \el{12}{C}, \el{14}{N}, \el{16}{O} and \el{20}{Ne}. A detailed description of the code and its features can be found in \citet{2006MNRAS.370.1817S} and references therein.

\begin{figure}
\includegraphics[width=\columnwidth]{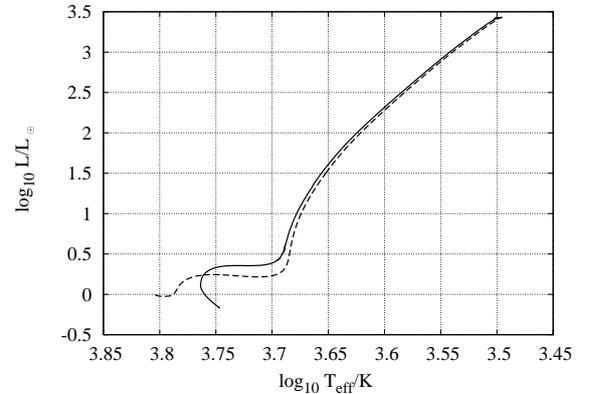}
\caption{HR diagram showing the evolution of a regular 1\ms\ model (solid line) and our standard 1\ms\ model which does not include the pp-burning reactions. The standard model is hotter because the star must contract more before the CNO elements settle to their equilibrium value.}
\label{fig:HRstandard}
\end{figure}

All our experiments have been performed on a 1\ms\ model of metallicity Z=0.02. In our standard run we do not consider the pp-chains. The reason for this is that in some of the experiments the use of the pp-chains leads to models that do not have giant-like composition profiles: they are not shell-like. The hydrogen abundance declines slowly towards the interior, rather than having a sharp drop at a particular location. By using only the CNO-cycle reactions to burn hydrogen, we obtain much sharper, shell-like profiles. The use of only the CNO-cycle reactions presents us with a problem: the more concentrated energy release leads to the formation of a convective core. As we want the star to behave in a similar way to a normal 1\ms\ model, we prevent any mixing from occurring in the convective core\footnote{It would also be desirable to force the code to use the radiative temperature gradient, essentially removing convection completely from the core. Attempts to do this proved unsuccessful as the models failed to converge.}. Figure~\ref{fig:HRstandard} shows a Hertzsprung-Russell (HR) diagram for our standard, CNO-burning only model and a normal 1\ms\ model that includes both the pp-chains and the CNO-cycle.

\subsection{Suppression of surface convection}
\begin{figure}
\includegraphics[width=\columnwidth]{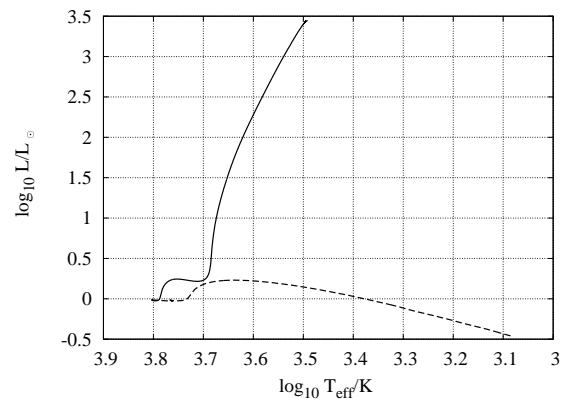}
\caption{HR diagram showing the evolution of the model without surface convection (dashed line). The standard model is also displayed for reference (solid line).}
\label{fig:HRnosurfaceconv}
\end{figure}

In this model, we suppress the occurrence of surface convection by forcing the code to use the radiative temperature gradient and preventing any mixing of the chemical elements. The resulting evolutionary track is displayed in Figure~\ref{fig:HRnosurfaceconv}. The starting model was taken at the ZAMS of the standard model. The model makes a sudden transition to lower temperatures at the beginning of the sequence because the star has to recover from the sudden loss of its surface convection zone.

Rather than evolving to higher luminosities after the end of the main sequence, this model proceeds to lower surface temperatures. The radiative envelope does not transport energy efficiently like a convective envelope does. Instead, the radiation is absorbed and the star's radius increases, pushing the star to lower surface temperatures. This star still becomes a giant, though it does not reach as high a luminosity as the standard model. The existence of surface convection is not important for a star becoming a red giant.

\subsection{Homogenous evolution}\label{sec:homogenous}
\begin{figure}
\includegraphics[width=\columnwidth]{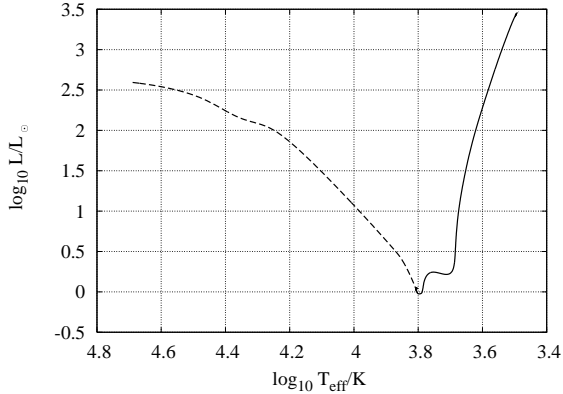}
\caption{HR diagram showing the evolution of the homogenous model (dashed line). The standard model is also displayed for reference (solid line).}
\label{fig:HRhomogenous}
\end{figure}

In this model, we force the whole star to be fully mixed throughout its evolution. The resulting evolutionary track is shown in Figure~\ref{fig:HRhomogenous}. In this model, the star does not become a giant -- its radius remains around one solar radius until hydrogen has been exhausted and then it begins to fall. The star moves blueward, not redward, throughout its evolution.

\subsection{The role of the mean molecular weight}
The homogenous model suggests that the mean molecular weight may play a role in a star's journey to gianthood. To investigate this, we have made a set of models that do not convert hydrogen to helium in the normal way. Instead, we burn hydrogen into a false element which we shall call pseudohelium. This pseudohelium is treated like normal helium in all respects but one. While it contributes to the opacity in the same way as helium and has the same ionization states and number of electrons as helium, we assign it a different atomic mass. By running various model sequences with different atomic masses for the pseudohelium, we can examine the effect that the mean molecular weight has on whether a star becomes a red giant.

\begin{figure}
\includegraphics[width=\columnwidth]{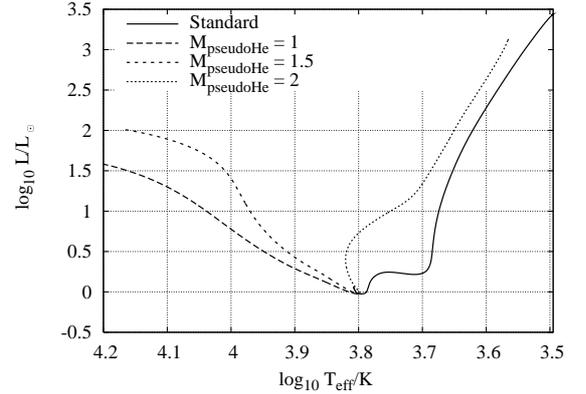}
\caption{HR diagram showing the evolution of the models with pseudohelium of varying atomic masses. The standard case is displayed for comparison and corresponds to $M_\mathrm{pseudoHe}=4$.}
\label{fig:HRpseudoHe}
\end{figure}

We have run test cases where the mass of the pseudohelium is 1, 1.5 and 2. The results are shown in Figure~\ref{fig:HRpseudoHe}. It should be noted that the case $M_\mathrm{pseudoHe}=1.5$ keeps the mean molecular weight of the star constant (assuming full ionization) because the pseudohelium has two electrons associated with it. If the atomic mass of the pseudohelium is below 1.5, the star does not become a giant nor does it become red. For a pseudohelium mass of 2, the star does evolve into a red giant with a compact core and an extended envelope. Experimenting with pseudohelium  masses between 1.5 and 2 suggests that this transition is smooth.

This sequence of tests, together with the earlier homogenous model, suggests that the mean molecular weight has a key role to play. If there is an insufficient change in the mean molecular weight throughout the star, the star does not become a giant. We shall return to this point later.  

\subsection{The role of energy generation}
We should not expect a change in the mean molecular weight to be solely responsible for a star's transition to giant status. This is evident when one considers the transition a star makes from the giant branch to the horizontal branch. The burning of helium to carbon and oxygen raises the mean molecular weight (even more so than the conversion of hydrogen to helium) yet the envelope shrinks. The nuclear burning also increases the energy generation in the core. One may therefore ask the question ``What role does energy generation play in a star's transition to a giant?''. To investigate this, we make the following tests.

\subsubsection{Composition-independent energy generation}
In this sequence, we allow hydrogen to be converted to helium in the usual way. However, we do not allow this process to generate energy. Instead the energy generation is given by
\be
\epsilon = \epsilon(\rho, T, X_\mathrm{H}=0.7)
\ee
i.e. we generate energy using the current temperature and density of the model, but assume that the hydrogen abundance is the ZAMS value. The energy generation is thus decoupled from the chemical evolution: hereafter we refer to this model as the `energy decoupled' case. The energy generation of the model remains centrally concetrated throughout the evolution. 

\begin{figure}
\includegraphics[width=\columnwidth]{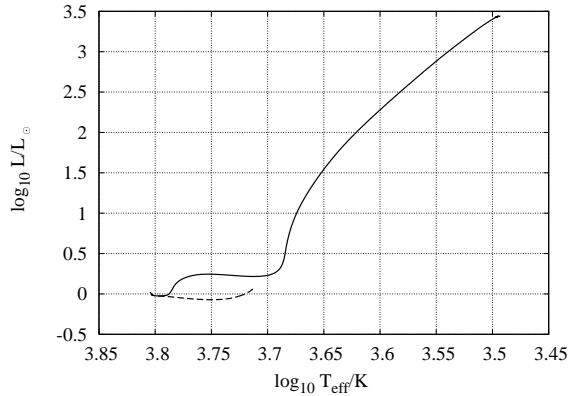}
\caption{HR diagram showing the evolution of the energy decoupled model (dashed line). The standard model is shown for comparison (solid line).}
\label{fig:HRdecoupled}
\end{figure}

The evolution of this model is shown in Figure~\ref{fig:HRdecoupled}. This initially looks promising: the model moves redward and we find that its radius is indeed increasing. However, closer inspection of the model shows that we have not made a giant in the sense that we wish. If we plot the radius profiles of this model, we find that all layers are expanding as the model evolves. This does not give a structure that has a compact core with an extended envelope -- our definition of a giant. This star is more akin to a star on its pre-main sequence.

\subsubsection{Forced profile}\label{sec:forced}
\begin{figure}
\includegraphics[width=\columnwidth]{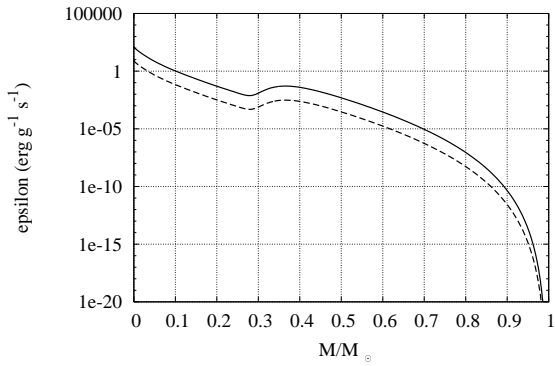}
\caption{The energy generation (solid line) and neutrino loss (dashed line) profiles as a function of mass adopted for the `Forced Profile' run.}
\label{fig:epsilon_forced}
\end{figure}

The energy decoupled model has one unintended feature: the energy generation rate (and its profile as a function of mass) changes with time. It is desirable to separate out these changes from the chemical ones. To do this, we take the energy generation profile (and the neutrino loss rate) as a function of mass at the beginning of the main sequence and impose this profile throughout the rest of the evolution. These profiles are displayed in Figure~\ref{fig:epsilon_forced}. The chemical evolution is allowed to occur as the usual function of temperature, density and composition.

\begin{figure}
\includegraphics[width=\columnwidth]{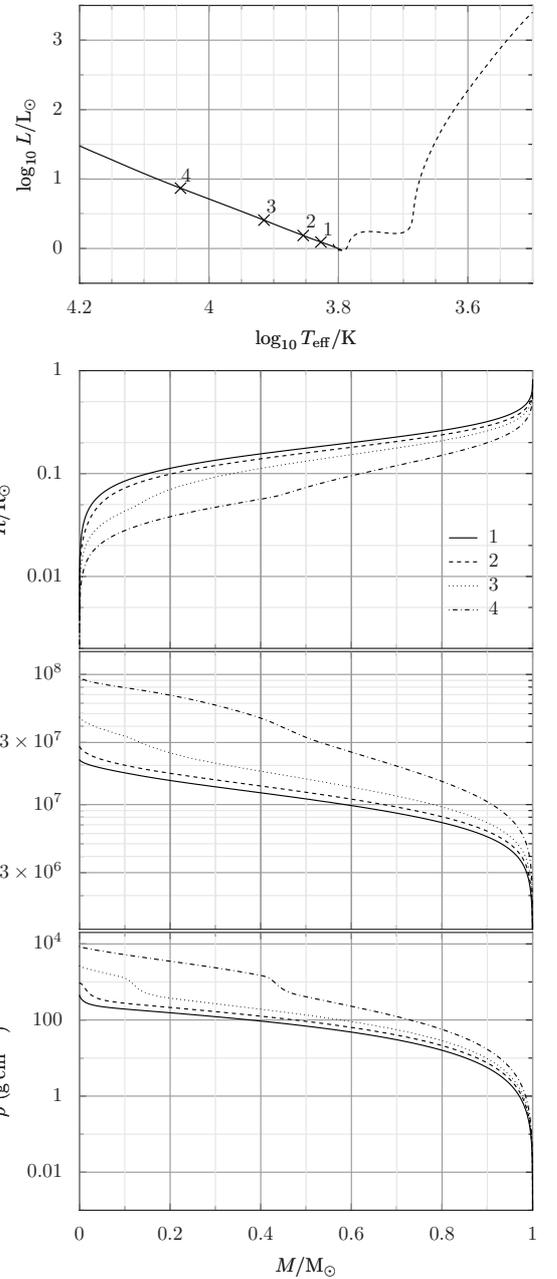}
\caption{Top panel: HR diagram showing the evolutionary track of the model with the energy generation profile fixed to its ZAMS value (solid line). The standard model is shown for comparison (dashed line). Crosses mark the points in the evolution at which the profiles in the lower panel are taken. Lower panel: Radius, temperature and density profiles for the model at the points denoted by crosses in the upper panel.}
\label{fig:HRForced}
\end{figure}

The evolution of this model in the HR diagram is shown in the top panel of Figure~\ref{fig:HRForced}. The model does not become a giant. It evolves to the blue, not the red, with all layers of the star contracting. The model has been evolved up to the point where the inner 0.7\ms\ has become hydrogen-exhausted. The temperature, density and radius profiles for this model show a distinct core has been formed (Figure~\ref{fig:HRForced}, lower panel). The increase in the surface luminosity of this model is due to the release of gravitational energy as the star contracts.

\section{Discussion}
Of the above tests, two things stand out. First, the homogenous model and the pseudohelium model which keeps the mean molecular weight constant do not become giants. Both these models evolve blueward with little change in their radii. This is despite their hydrogen-burning luminosities increasing by over two orders of magnitude from the ZAMS value. We may conclude that the mean molecular weight plays some role in whether a star becomes a red giant or not. Secondly, the forced profile model also does not become a giant, despite it developing a strong mean molecular weight gradient and forming a core-like structure (in terms of composition -- the density remains somewhat lower than in the standard case). We therefore conclude that a mean molecular weight gradient {\it alone} is not sufficient to determine whether a star becomes a red giant or not, but it is necessary.

So what other conditions must be met if we are to get a red giant? One obvious candidate is the rate of energy production. Our energy decoupled model shows that we can get redward motion if sufficient energy is deposited into the star. This should seem reasonable: in order to expand a gas, work has to be done and hence an energy source is needed.

To investigate the role that energy generation may play in whether a star becomes a red giant or not, we make the following tests. Firstly, we start from a model which has a well-defined mean molecular weight gradient. We have chosen to use a model from the forced profile sequence, taken at a point when the mass of the hydrogen-exhausted core is 0.2\ms. We also stop hydrogen from being converted into helium, as we are trying to determine how the model reacts to changes in the energy generation only. Any changes in the chemical structure could potentially be confusing. Because of this, the models we obtain should not be regarded as `evolutionary' models -- they are merely the route the star takes in relaxing from one set of conditions to another. With these modifications, we are now free to alter the energy generation profile from that used in the forced profile sequence.

The first change we make is to simply increase the energy generation profile by a fixed amount by multiplying it by some factor. We also do the same to the neutrino loss rate profile. If we multiply these profiles of a factor of between 2 and 5, the star contracts and moves blueward. If the energy generation profile is multiplied by a factor of ten, we obtain different behaviour. The star expands and moves redward. However, all the layers of the star are expanding so we do not get the desired compact core/extended envelope structure of a giant.

One may then ask whether the way in which we deposit energy into the star is important. In a real red giant, energy is released in a shell rather than throughout the star as the above models have assumed. We therefore attempt the following. We take the energy generation profile from the `forced profile' sequence and the same input model as above. This time, instead of multiplying the whole energy generation profile, we increase the energy generation rate in a well defined region. This is done by adding a narrow Gaussian term at a given mass. We have tried placing the term at 0.1, 0.3 and 0.5\ms, which correspond to additional energy generation inside the hydrogen-exhausted region (i.e. the `shell') of the initial model, just outside the H-exhausted region, and within the envelope repsectively.

\begin{figure}
\includegraphics[width=\columnwidth]{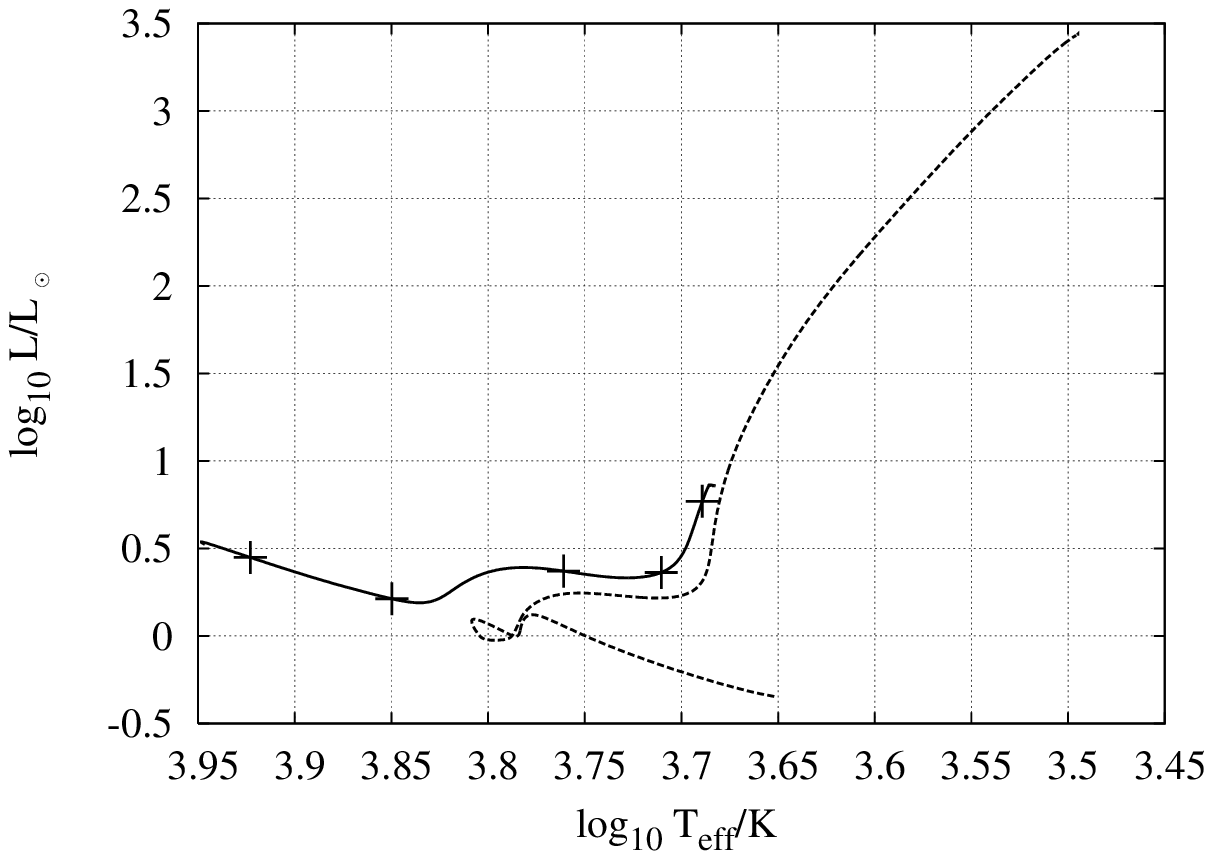}
\includegraphics[width=\columnwidth]{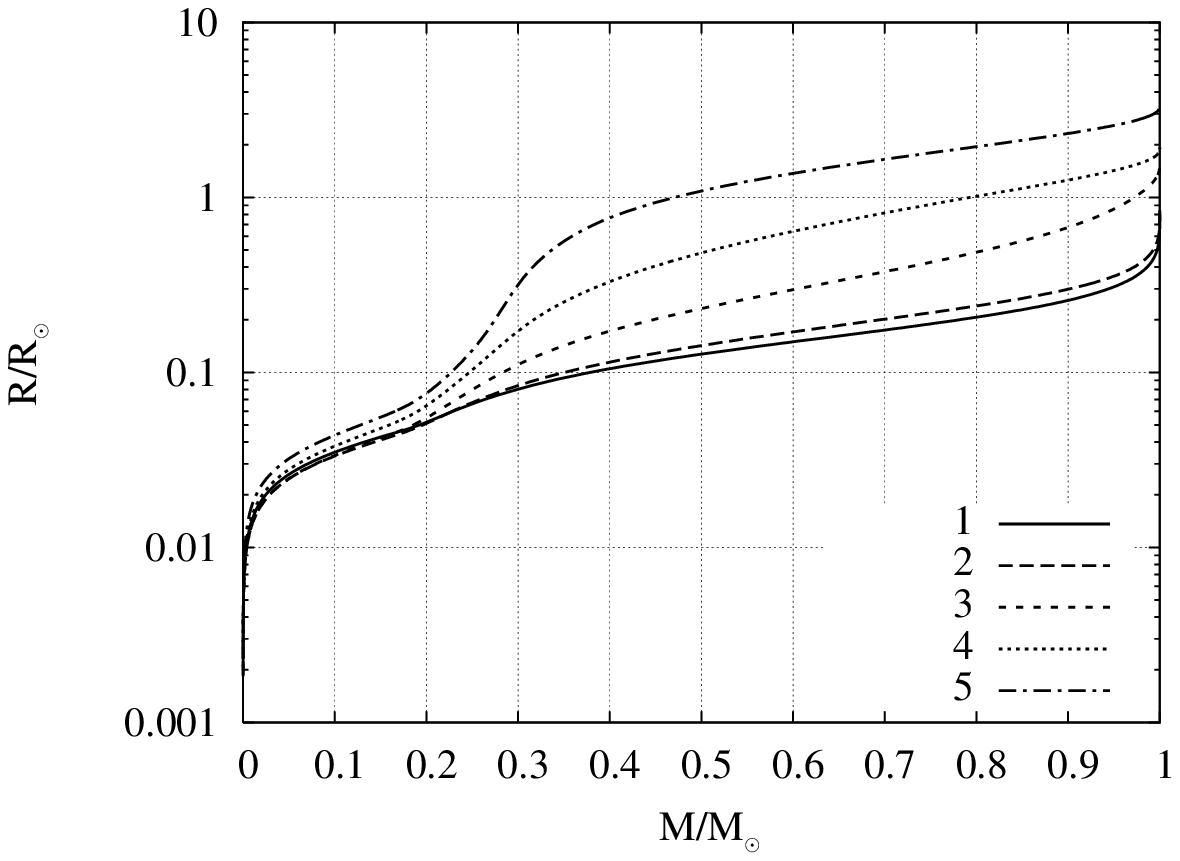}
\caption{Top panel: HR diagram showing the track of the model where energy is injected at 0.3\ms\ (solid line), with the evolution being redward. The standard track is shown for comparison (dashed line). The crosses denote points at which the detailed profiles in the lower panel have been taken. Bottom panel: Radius as a function of mass for selected models in the sequence. Note that significant expansion occurs only above 0.2\ms.}
\label{fig:Gaussian}
\end{figure}

The results for the case of energy injection at 0.3\ms\ are shown in Figure~\ref{fig:Gaussian}. We plot the evolution of the radius profile for several consecutive models. We see that above 0.2\ms\ there is a substantial increase in the radius for a give mass. Crucially, we have expansion {\it inside} the region at which energy is injected and it is only within the H-exhausted region that little change in radius takes place. We obtain similar results if the energy is injected at 0.5\ms.

As discussed above, it appears that a mean molecular weight difference is necessary for a star to become a giant and we can illustrate this with the following test. We select a model without a strong molecular weight gradient from the forced model sequence and inject energy using the same Gaussian profile as above. We have chosen a model about halfway through the main sequence. In this case, we find that the entire star expands {\it regardless} of where we inject the energy. Thus we deem the change in mean molecular weight to be a necessary condition for forming a giant.

\subsection{A spanner in the works: mass dependence}
\begin{figure}
\includegraphics[width=\columnwidth]{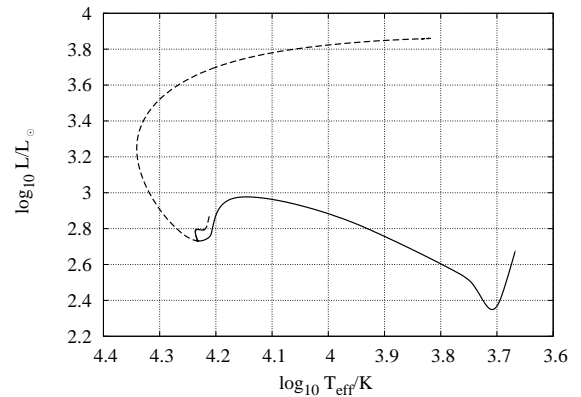}
\caption{HR diagram showing the evolution of a 5\ms\ model evolved with a pseudohelium atomic mass of 1.5 (dashed line). A standard 5\ms\ model is shown for comparison (solid). Both models use only CNO burning reactions and no mixing has been permitted in convective regions. The standard evolution has been terminated on the giant branch.}
\label{fig:HR_5}
\end{figure}

The model of \citet{1993ApJ...415..767I} which has constant opacity and constant mean molecular weight becomes a giant. This is clearly at odds with the picture presented above, in which the mean molecular weight gradient plays a key role. We note that Iben's model is for a star more massive than our model (his model is 5\ms) so it seems sensible to repeat our pseudohelium experiment for this mass of star. It is here that we run into a problem. The evolution of this model is shown in Figure~\ref{fig:HR_5}, along with a standard\footnote{That is, one that also uses only CNO burning, has no convective mixing and which burns hydrogen to normal helium in the same way that our standard 1\ms\ model did. This is not a normal 5\ms\ model!} 5\ms\ model. Our 5\ms\ model, like that of Iben, does indeed become a giant even in the case where we keep its mean molecular weight constant (i.e. by setting the atomic mass of the pseudohelium to 1.5). There is only one conclusion we can drawn for this test -- the mean molecular weight is not crucial to a star becoming a giant {\it in all masses of star}. However, we note that if we evolve a 5\ms\ model in a homogenous manner (i.e. ensuring that the whole star is mixed as we did in section~\ref{sec:homogenous}), it contracts just as the 1\ms\ model did.

One of our critical tests in the case of the 1\ms\ model was the one in which we forced the energy generation profile to remain as it was on the main sequence (the `forced profile' sequence of section~\ref{sec:forced}). This demonstrated that the mean molecular weight gradient alone was not enough to produce a giant. We have re-run this test for the 5\ms\ model, taking the zero-age main sequence energy generation profile for this mass of star. We find that this star does not become a giant, even though we are able to grow an extremely large core (over 80\% of the star's mass). It seems that an increase in the stars luminosity is indeed a necessary condition for a star to become a red giant. This is perhaps an unsuprising result: in order to expand something, work must be done and this requires an input of energy.

We are still in the process of investigating the 5\ms\ model in order to determine why it becomes a giant. However, the fact that a more massive star can become a giant when it has a constant mean molecular weight is extremely important. Most work on why stars become red giants is done under one fundamental assumption, namely that there is only one way to make a giant\footnote{The assumption may be correct: it may appear that there is more than one way to make a giant simply because we are ignorant of the factors that cause it.} and it is common to all stars, regardless of mass (and perhaps other properties too, such as metallicity). As such, many studies have only examined one particular mass of star \citep[e.g.][]{1992ApJ...400..280R,1993ApJ...415..767I} and one should therefore take their conclusions in that light.

That different masses of star may behave differently is hinted at by \citet{1998MNRAS.298..831E}. Using polytropic models, these authors demonstrated that a jump in the mean molecular weight could affect  the Sch\"onberg-Chandrasekher limit. They found that if the ratio of the core mean molecular weight to that of the envelope is less than 3 then the core mass can be arbitrarily large, though there is a limit to the core radius. We may speculate that our 1 and 5\ms\ models behave in different ways because they sit on different sides of this (or a similar) threshold. The 1\ms\ model may not become a giant if it does not have a mean molecular weight gradient because it is unable to form a core. This connection will be looked at in future work. In addition, the tests we have carried out should be repeated on a range of masses to see if there are further difference we have not yet noticed and which might yield clues as to why stars become red giants.

\section{Conclusions}
We have experimented with the inputs to a stellar model in order to work out why stars become red giants. Our models show two things seem to be important in our 1\ms\ star:
\begin{itemize}
\item{A star must have a mean molecular weight gradient if it is to become a giant, but this is not a sufficient condition for it to do so.}
\item{Sufficient energy must be supplied in order for the star to become a giant.}
\end{itemize}
The mean molecular weight is merely a tool -- the star must have enough power to make use of that tool.

However, this cannot be the whole story as to why stars become red giants. Running the same battery of tests on a 5\ms\ model yields a different result. Our 5\ms\ star can become a giant {\it without} a mean molecular weight gradient. However, sufficient energy must still be supplied to the star in order for it to expand just as in the case of our 1\ms\ model. We suggest that there may be more than one way of making a giant and we will address this problem in future work.

\section*{Acknowledgements}
The authors gratefully thank the referee, Noam Soker, for his stimulating comments which have both improved this manuscript and provided fodder for a forthcoming follow-up investigation. RJS and RPC are funded by the Australian Research Council's Discovery Projects scheme under grants DP0879472 and DP0663447 respectively.

\bibliographystyle{apj}
\bibliography{../../../masterbibliography}

\end{document}